\documentclass[11pt,twoside]{article}
\usepackage{asp2006}
\usepackage{epsf}
\usepackage{psfig}
\usepackage{graphics}
\usepackage{lscape}
\usepackage{natbib}

\begin{document}

\title{An Improved Black Hole Mass--Bulge Luminosity Relationship for AGNs}

\author{C. Martin Gaskell and John Kormendy}
\affil{Astronomy Department, University of Texas, Austin, TX 78712-0259, USA}

\begin{abstract}
Two effects have substantially increased the scatter in the AGN black hole mass--host galaxy
bulge luminosity relationship derived from SDSS spectra.  The first is that at a fixed black hole mass,
$M_\bullet$, the SDSS spectrum depends strongly on redshift because an SDSS fiber sees a larger fraction of the total
light of more distant galaxies. The second is that at a given redshift, the fraction of host-galaxy light in the
fiber increases with decreasing galaxy luminosity. We illustrate the latter effect using the Kormendy et al. (2009)
light profiles of Virgo ellipticals. With allowance for the two effects, we obtain a black hole mass---bulge
luminosity ($M_\bullet$ -- $L_{host}$) relationship for AGNs which has a scatter of only $\pm 0.23$ dex in mass.
This is less than the scatter found for inactive galaxies, and is consistent with the measuring errors.  We show that there is a
corresponding tight linear relationship between the fraction of host galaxy light in AGN spectra, $L_{host}$/$L_{AGN}$,
and the Eddington ratio, $L/L_{Edd}$.  This linearity implies that at a given $M_\bullet$, host luminosities of
high-accretion-rate AGNs (NLS1s) and low-accretion-rate AGNs are similar. The $L_{host}$/$L_{AGN}$ -- $L/L_{Edd}$
relationship provides a simple means of estimating the fraction of host galaxy light in AGN spectra.  This means that
the real amplitude of variability of low-accretion-rate AGNs is increased relative to NLS1s.

\end{abstract}

\section{Introduction}
It has long been recognized
that the masses, $M_\bullet$, of supermassive black holes (SMBHs) are proportional to the stellar luminosities, $L_{host}$, of the bulges of the galaxies in which they are located (see \citealt{kormendy+richstone95} and \citealt{kormendy+gebhardt01} for reviews).  $M_\bullet$ can be most easily determined for AGNs. \citet{dibai77} showed that $M_\bullet$ can be estimated
for an AGN from the broad emission lines in a single-epoch spectrum, so long as the broad-line region (BLR) motions are gravitationally dominated. Reverberation mapping has verified that the BLR motions are dominated by gravity \citep{gaskell88} and has confirmed the
accuracy of the Dibai method (see \citealt{bochkarev+gaskell09}).  With the advent of the SDSS, the Dibai method has been used to estimate the masses of tens of thousands of black holes. \citet{shen+08} give masses for 900 SDSS AGNs for which
\citet{vanden_berk+06} have spectroscopically estimated $L_{AGN}/L_{host}$, the ratio of AGN light to host galaxy light at 5100 \AA\ in the rest frame.  In this paper we use these data to study the dependence of $L_{AGN}/L_{host}$ on the accretion rate (which we will express by the Eddington ratio, $L/L_{Edd}$) and to obtain an improved  $M_\bullet$ -- $L_{host}$ relationship.

\begin{figure}[t!]
\begin{center}
\plottwo{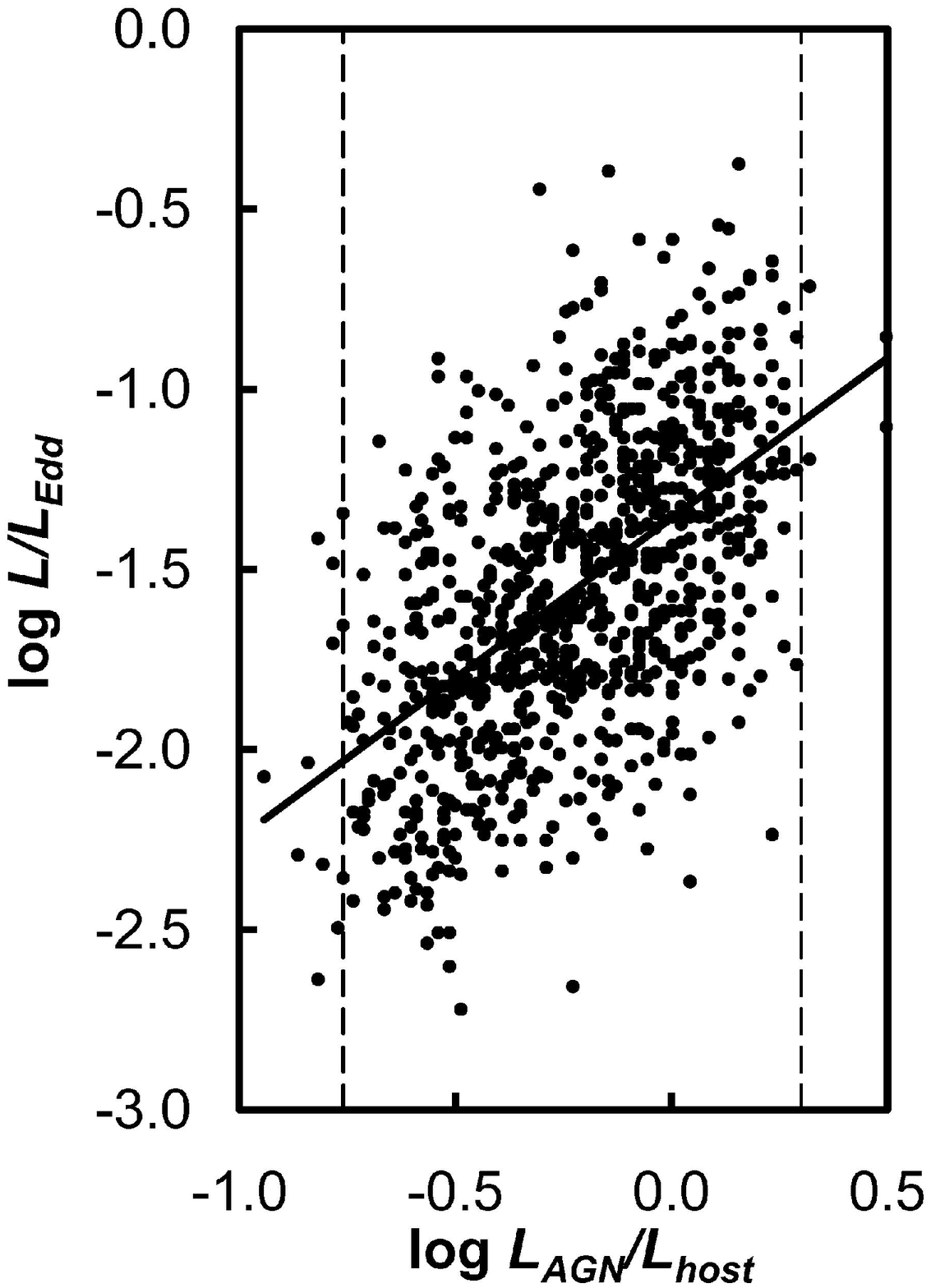}{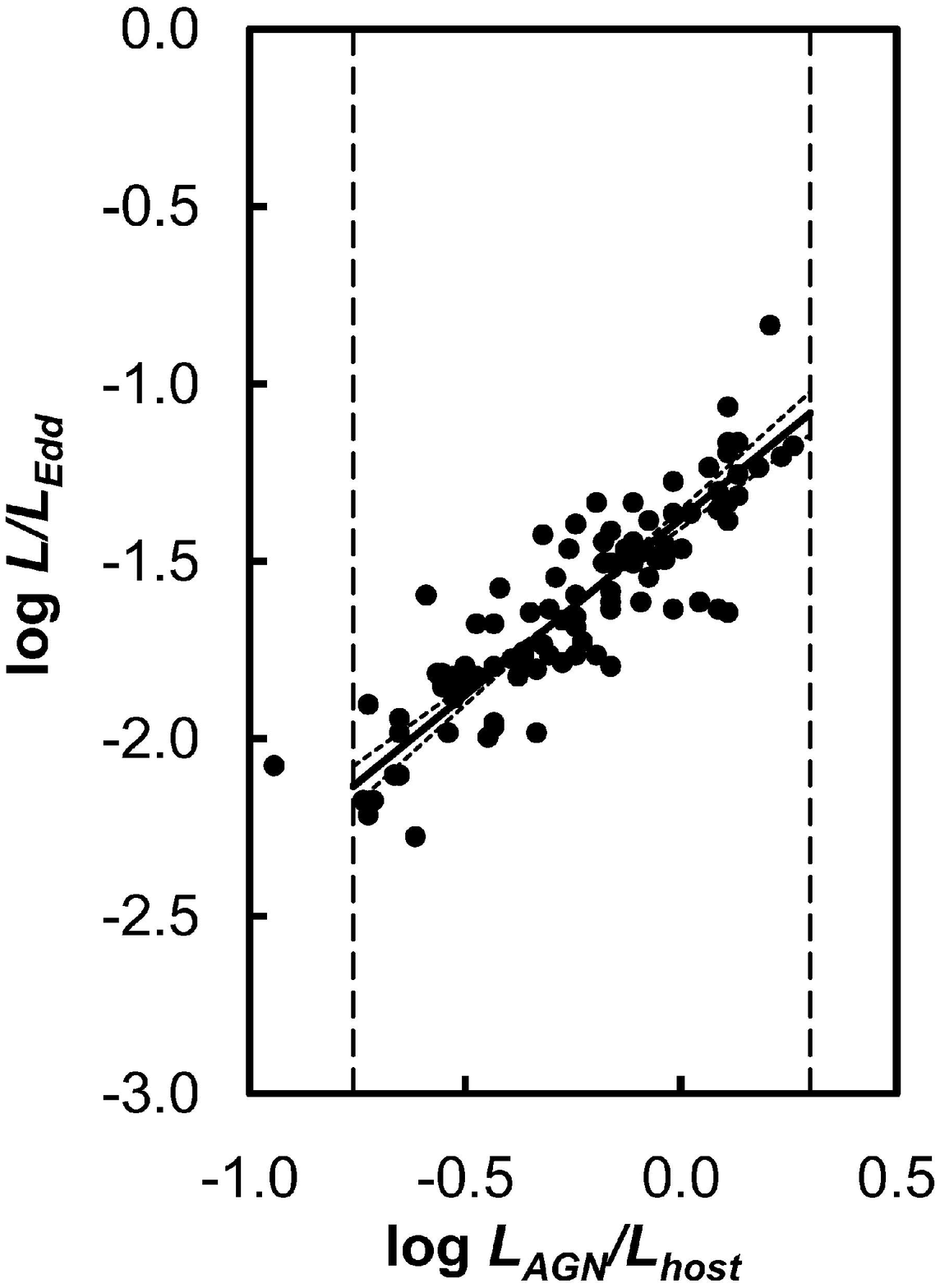}
\caption{Eddington ratios, $L/L_{Edd}$, as a function of $L_{AGN}/L_{host}$, the ratio
of AGN flux to host galaxy flux at 5100 \AA.  The dashed
vertical lines show the approximate upper and lower cutoffs of \citet{shen+08}.  (a) (left panel) shows all low-redshift ($z < 0.45$)
SDSS AGNs in the sample. The diagonal line is a linear regression on $\log L_{AGN}/L_{host}$. (b) (right panel) shows just 100 AGNs with $\log M_{bh} = 7.7 \pm 0.2$, and
$0.13 < z < 0.18$.  The solid diagonal line is a censored OLS-bisector fit \citep{isobe+90}, and the 68\% confidence interval for the slope is shown
by the two dotted lines.
\vspace{-0.5cm}}
\end{center}
\end{figure}

\section{Results}

Fig.\@ 1a shows $L_{AGN}/L_{Edd}$ as a function of $L_{AGN}/L_{host}$.  Because of difficulties in measuring $L_{AGN}/L_{host}$
when the AGN is too bright or too faint compared with the galaxy, upper and lower cutoffs on the ratio have been imposed by \citet{shen+08}.  Since $L_{AGN}$ appears on both axes and since $M_\bullet \sim L_{host}$, we expect a simple linear correlation between $L/L_{Edd}$ and $L_{AGN}/L_{host}$.  Fig.\@ 1a does indeed show a correlation, but the scatter is very large.  This scatter could be a consequence of measurement errors, or it could reflect intrinsic scatter in the $M_\bullet$ -- $L_{host}$ relationship.

In Fig.\@ 2a we show that if we take a narrow range of $M_\bullet$, then the luminosity deviation, $\Delta
\log L_{host}$, from the diagonal line in the left panel of Fig.\@ 1 is a strong function of redshift.  This has a simple explanation: at low redshift an SDSS fiber is only taking in a small part of the bulge of the host galaxy, so the luminosity of the bulge is underestimated for nearby galaxies.  Fig.\@ 2b uses the \citet{kormendy+09} photometry of Virgo ellipticals to show that at a given redshift, the fraction of total bulge light in a an SDSS fiber decreases as the luminosity of the galaxy increases.   The combination of the two effects will be discussed in detail elsewhere (Gaskell \& Kormendy, in preparation), but the dramatic improvement in the $L/L_{Edd}$ -- $L_{AGN}/L_{host}$ relationship can be illustrated simply by plotting AGNs in a narrow range of $z$ and black hole mass.  This is shown in Fig.\@ 1b.

\begin{figure}[t!]
\begin{center}
\plottwo{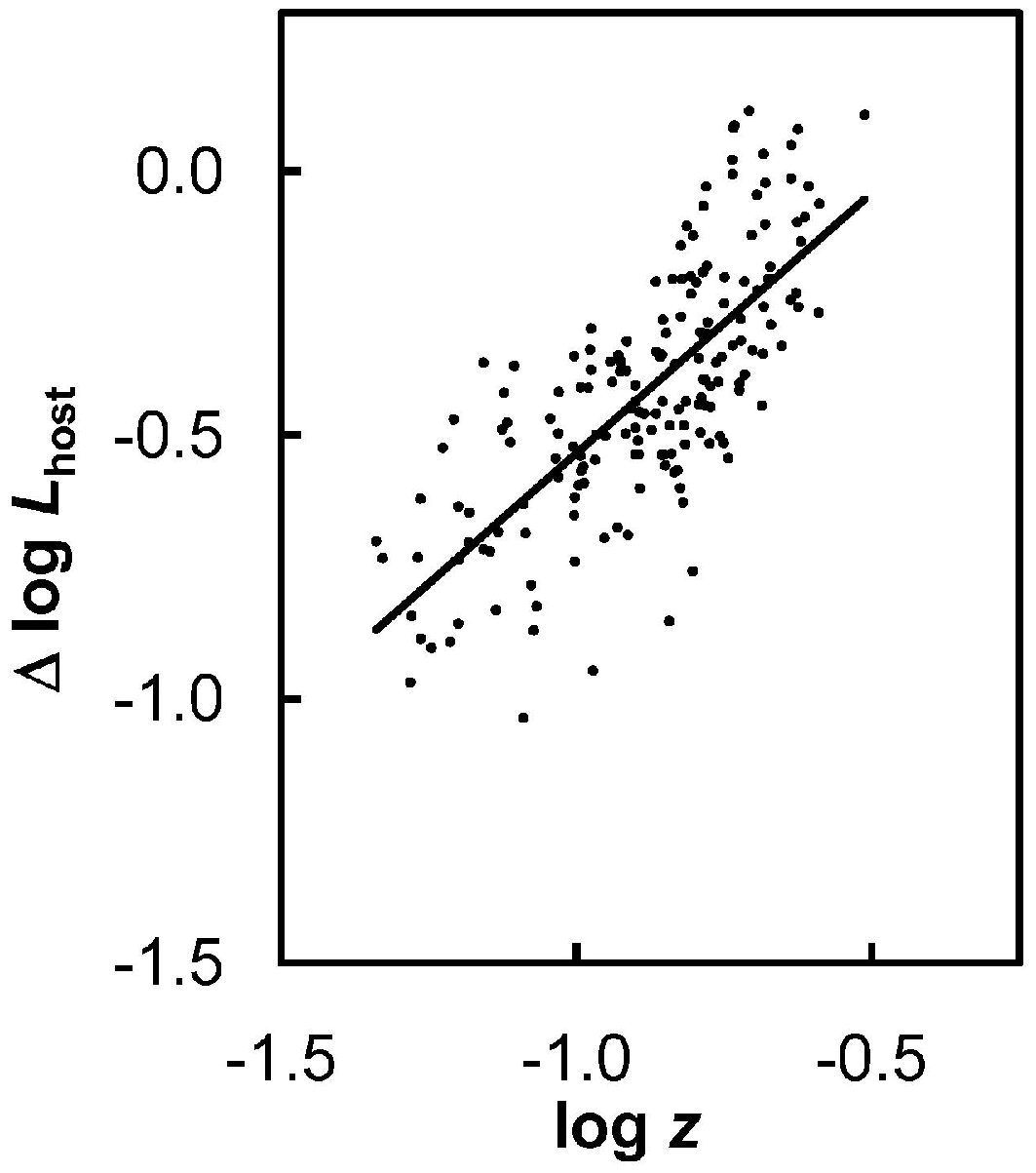}{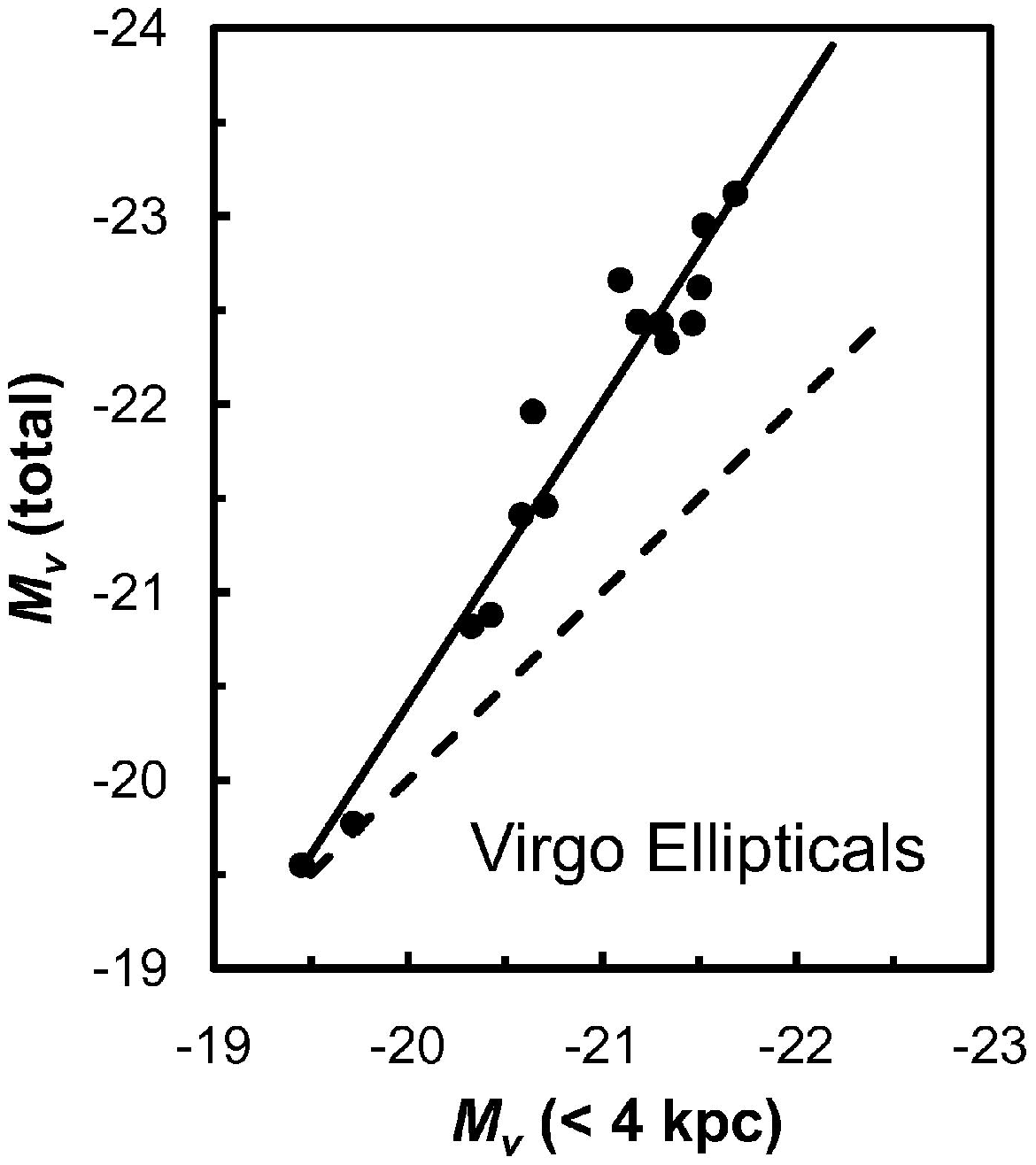}
\caption{(a) estimated relative deficit, $\Delta \log L_{host}$, of host galaxy light at $\lambda$5100 as a function
of redshift for AGNs with $7.4 < \log M_\bullet < 7.6$.  $\Delta \log L_{host}$ is
normalized to $\log z = -0.4$. The diagonal line is a linear regression on $\log z$. (b) The total absolute magnitudes of Virgo ellipticals (derived from \citealt{kormendy+09}) as a function of the absolute magnitude measured within
a fixed 4 kpc radius aperture.  The solid line is a linear regression of the total magnitude
on the 4 kpc magnitude.  The dashed line shows what the relationship would be if all the light were within 4 kpc.
\label{all_data}
\vspace{-0.5cm}}
\end{center}
\end{figure}

Fig.\@ 3 shows the resulting $M_\bullet$ -- $L_{host}$ relationship for AGNs over a narrow redshift range.  $L_{host}$ has been approximately
corrected using the relationship for the Virgo ellipticals in Fig.\@ 2b.  The dispersion in mass in Fig.\@ 3 is $\pm 0.23$ dex which is better than the $\pm 0.30$ dex dispersion \citet{haring+rix04} found for the $M_\bullet$ -- $L_{host}$ relationship for non-active galaxies.
Applying the luminosity correction from Fig.\@ 2b also increases the slope of $M_\bullet$ -- $L_{host}$ relationship for the SDSS AGNs.  If we take an $M_\bullet$ -- $L_{host}$ relationship of the form $M_\bullet \propto L_{host}^\alpha$, then for the complete uncorrected \citet{shen+08} sample (not shown), an OLS-bisector fit gives a slope of  $\alpha = 0.69 \pm 0.02$, while a similar fit for the corrected subset of 100 AGNs in Fig.\@ 3 gives $\alpha = 0.84 \pm 0.03$.  If we take the luminosity-dependence of the mass/light ratio of bulges to be $M/L \propto L^{0.32 \pm 0.04}$ \citep{cappellari+06}, then $M_{bulge} \propto M_\bullet^{1.16 \pm 0.07}$.

\begin{figure}[t!]
\begin{center}
\epsfxsize = 70 mm
\epsfbox{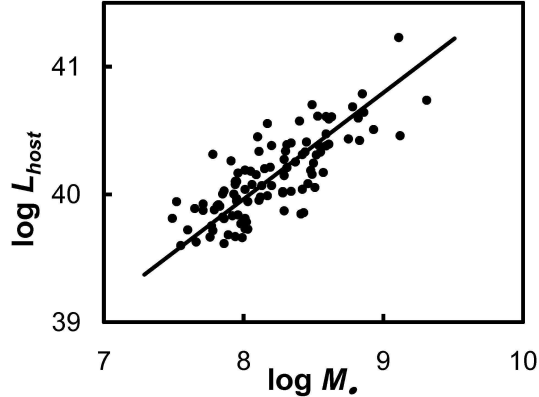}
\caption{The $M_\bullet$ -- $L_{host}$ relationship for
100 AGNs restricted to $0.13 < z < 0.34$ and with $L_{host}$ corrected for the aperture effect. The diagonal line is the OLS-bisector fit,
$M_\bullet \propto L^{0.84}_{host}$.
\label{all_data}
\vspace{-0.5cm}}
\end{center}
\end{figure}

\section{Discussion}

The small dispersion in the AGN $M_\bullet$ -- $L_{host}$ relationship in Fig.\@ 3 implies that the intrinsic relationship must be very tight (as tight as the $M_\bullet$ -- $\sigma_*$ relationship), since much or all of the scatter can be accounted for by observational errors.
The small dispersion also supports the reliability of AGN black hole masses determined via the Dibai method.  As discussed in \citet{bochkarev+gaskell09},
this implies that for the AGNs for which the method has been used, both the structure and intrinsic spectral
energy distributions are very similar \citep{gaskell+04,gaskell+benker09}.  The slope  in Fig.\@ 1b ($0.99 \pm 0.11$) shows that the hosts of high-accretion-rate AGNs do not systematically deviate from the $M_\bullet$ -- $L_{host}$ relationship.

Fig.\@ 1b makes determination of host-galaxy contamination of AGN photometry and spectroscopy
straight forward.  One immediate result is to show that low $L/L_{Edd}$ AGNs must have greater optical variability amplitudes
on average than high $L/L_{Edd}$ AGNs (NLS1s).  \citet{klimek+04} have already shown that even without corrections for host galaxy
contamination, NLS1s seem to be less variable than non-NLS1s.  Fig.\@ 1b shows that this difference will be much greater when
the higher galaxy contamination of non-NLS1s is allowed for.

\begin{acknowledgments}
This research has been supported through NSF grants AST 08-03883 (MG) and AST06-07490 (JK).

\end{acknowledgments}

\end{document}